# D2D Multi-hop Energy Efficiency Toward EMS in B5G

Asma AL-Mansor
*Department of Computer and Communication System Engineering, Universiti Putra Malaysia, 43400UPM Serdang, Malaysia*
asmahussein282@gmail.com

Nor Kamariah Noordin
*Department of Computer and Communication System Engineering, Universiti Putra Malaysia, 43400UPM Serdang, Malaysia*
nknordin@upm.edu.my

Abdu Saif
*Department of Electrical Engineering, Faculty of Engineering, University Malaya, 50603, Kuala Lumpur, Malaysia*
saif.abduh2016@gmail.com

Saeed Hamood Alsamhi
*NUIG, Galway, Ireland, Faculty of Engineering, IBB University, Ibb, Yemen*
saeedalsamhi@gmail.com

*Abstract*—Beyond fifth generation (B5G), communication has attracted much attention from academia, industry, and mobile network operators due to network densification, ultra-low latency communication, and enhanced energy and spectrum efficiency. However, a post-disaster emergency management system (EMS), which increasingly relies heavily on wireless communication infrastructure, is falling far behind in innovation and funding. Because the B5G concept represents a telecommunications industry revolution, EMS provisioning is intended to be dispersed, autonomous, and robust to network weaknesses caused by human and natural calamities. When the network is congested, partially functioning, or entirely isolated, we provide multi- device-to-device (D2D) communication to extend the communication coverage area with improved energy efficiency. Furthermore, we examine D2D multi-hop energy efficiency performance in the proposed network. The results demonstrate that the improved D2D multi-hop energy efficiency can improve the EMS effectively and efficiently in extending the coverage area and enhancing energy efficiency. Moreover, the proposed approach has been proven to increase energy efficiency, which acts as a suitable network design to recover from natural disasters and potentially save many lives.

## I. INTRODUCTION

BEYOND fifth generation (B5G) offers ultra-low latency, high energy, and spectral efficiency, attracting much attention from academia, service providers, and industry. Device-to-device (D2D) communication is a promising technology for efficient connectivity to improve wireless network coverage. Multi-hop relay D2D communication assists the network in extending the coverage area and improving the system's performance. In this context, the authors of [1] discussed the coverage services through an increased number of relay hops and reducing the traffic load of the BS. D2D communications are also utilized in public safety networks (PSNs) to maximize system capacity and energy/spectrum efficiency [2]. The wireless infrastructure's status following a disaster depends on the kind of disaster, such as an earthquake, storm, or terrorist attack. Achieving compatibility across user devices and interoperability among post-disaster emergency management system (EMS) communication systems would necessitate a generic solution. This motivates stakeholders to create a disaster network solution that can handle as many disasters as feasible. Therefore, D2D is one of the critical solutions recommended for disaster recovery. As a result, users can communicate and connect to the nearest operable base stations (BS) directly or through nearby users by raising the uplink to transmit power. D2D communications are crucial for multi-hop communication because the first alternative is not optimum due to the power restriction. The establishment of multi-hop D2D communication requires a low-complexity control signal exchange protocol. Therefore, users should continually transmit and receive control signals to the BS and the nearest available D2D peers. However, it is worth noting that when a gadget processes a considerable number of control signals, its energy efficiency plummets. As a result, to allow the D2D mode for post-disaster communication, a new energy-efficient signaling protocol, resource allocation, and optimization approaches are required [3]. During natural disasters, the coverage service is critical to assisting accidental areas of disasters where infrastructure networks have been damaged or when wireless coverage to user devices is unavailable [4]. Many space technologies, such as satellites, high-altitude platforms, tethered balloons, and drones, are used to mitigate the impact of the disaster on economics and the environment [5-9]. The relay aids the BS network in achieving capacity over Rayleigh fading channels by providing multi-hop D2D communication. The author in [5] evaluates the best D2D communication design and the impact of the number of relay hops on system capacity and

TABLE I COMPARISON OF EXISTING WORKS

| Reference, Year | Highlights | 1 | 2 | 3 | 4 | 5 |
|---|---|---|---|---|---|---|
| [11], (2019) | A BS-assisted Wi-Fi network through a smartphone acts as a relay for its neighbor for rescue operations in disaster recovery. | ✓ | ✓ | ✗ | ✓ | ✓ |
| [12], (2022) | This is an overview of a strategy for dealing with any large-scale disasters through the recovery of terrestrial infrastructure by multi-hop communication between disaster and non-disaster areas | ✓ | ✓ | ✓ | ✓ | ✗ |
| [13], (2020) | The optimal location of BSs is examined to minimize the average path loss to the ground receiver. | ✓ | ✗ | ✗ | ✓ | ✓ |
| [14-16], (2020) | The tethered UAV is being used to support the cellular network for disaster recovery. | ✓ | ✓ | ✗ | ✗ | ✓ |
| [17], (2021) | The D2D, drone-assisted communication, and mobile ad hoc networks are proposed for post-disaster recovery in case the network is congested, partly functional or completely isolated. | ✓ | ✓ | ✓ | ✓ | ✗ |
| [18], (2018) | The BS is integrated with D2D communication to maximize the rate of a network system. | ✓ | ✗ | ✗ | ✓ | ✓ |
| [19], (2017) | A cellular network technique examines enhanced energy efficiency and spectral efficiency by multi-hop D2D communications for extended coverage. | ✓ | ✓ | ✓ | ✓ | ✗ |
| This work | We are focused on optimal relay-hops integrated with BSs to improve the coverage area, reduce the BS load, and minimize energy consumption for disaster areas. | ✓ | ✓ | ✓ | ✓ | ✓ |

**(1) BS  (2) optimal relay hops (3) D2D communication (4) post-disaster recovery  (5) coverage improvement**

power efficiency. In a natural disaster, BS and multi-hop D2D communications establish a link between user devices that are out of coverage [4, 10]. Therefore, extending BS coverage via multi-hop relay leads to improving the wireless.

The authors of [10] introduced disaster recovery options from the users' standpoint and network solutions like D2D and dynamic wireless networks. References in [20] addressed the topic of resilience, highlighting a variety of options like D2D, unmanned aerial vehicles (UAVs), and the Internet of things (IoT). In addition, IoT-enabled flood search-and-rescue systems are critically evaluated in [21, 22], and a novel IoT-aided integrated flood management framework based on water-ground-air networks is presented. Instead, where there are no operational cell towers, constructing multi-hop D2D connections is proposed to expand the coverage area of UAVs [4, 23, 24]. To reduce energy usage, the authors of [25] presented a hierarchical D2D design with a centralized software-defined networking controller that communicates with the cloud head.

Furthermore, the authors of [26-28] introduced using UAVs to find D2D devices in disaster areas. The relay will forward the wireless out of the BS's coverage, and users can get coverage service through multi-hop D2D communications [22]. We assume the BS configuration to centralize the transmission beam to an optimal relay for reliable connectivity.

The relay and D2D communications are utilized in a PSN to improve the capacity and spectrum efficiency while keeping the network's connectivity. In this context, in a PSN, communication recovery is critical for any natural disaster event to link dysfunctional and functional areas to save lives [29]. Therefore, BSs have limited recovery disaster communication, and thus an efficient system needs a long transmission distance for the larger area. An optimum relay is a promising solution to improving wireless coverage services by expanding coverage via multi-hop D2D communication and reliable connectivity in cellular network failure [30]. An e optimal relay node is selected based on the residual energy and link quality in the edge of BS coverage to increase the signal strength connectivity. The performance of an optimal relay extends range by growing hops and reliably provides wireless coverage services to remote user devices.

This paper proposes a system model for BS-assisted user devices in disaster areas in B5G networks by extending the coverage area and enabling D2D multi-hop energy efficiency. The contributions of this paper are summarized as follows:
1- We focus on integrated BSs with D2D multi-hop communication to deliver communication service, extend coverage, and improve service quality. The benefits of the proposed system model include providing coverage service to user devices in its coverage range, improving capacity, and enhancing energy efficiency.
2- We investigate the trade-off between increasing the number of relays and D2D multi-hop communication for extending the coverage regions with energy efficiency.
3- We analyze and discuss the energy efficiency of the hops for different distances and path loss exponents.

*A. Paper structure*

The remainder of the paper is organized as follows. Section II presents the system model. Section III offers the simulation of results for further analysis. Finally, Section IV concludes the paper.

II. SYSTEM MODEL

The system model is shown in Figure 1. In order to help public safety networks during emergencies, an active BS is utilized as a scenario. The user devices will be distributed using the Poisson cluster process in the disaster area. To relay wireless coverage to user devices in the disaster region, the edge of BS coverage user devices will be chosen as relay nodes. D2D communication actions help to expand system capacity and boost BS wireless range. We presume the statics are the BS locations and an ideal relay hop distance. $hop_1$ is considered between the BS and Relay, while $Hop_2$ is taken into account between Relay and Cluster Heads, and $hop_3$ is between Cluster Heads and CMs (D2D-link). The multi-hop

D2D communications and adaptive clustering architecture extend the BS coverage and improve energy and spectrum efficiency.

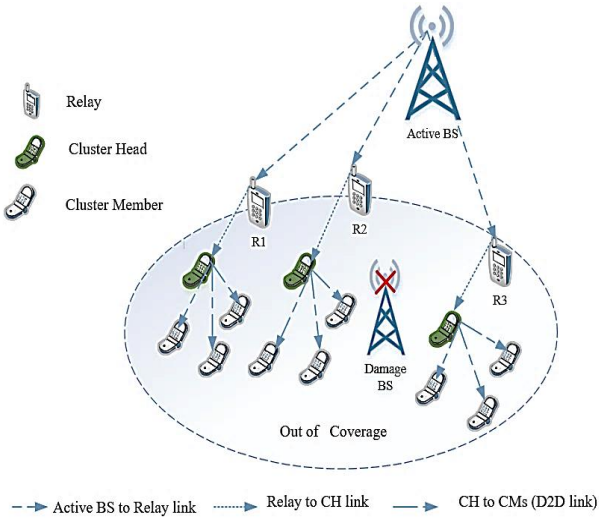

Figure 1. System model.

The BS provides the coverage to the user devices in full-duplex and relays nodes forward them through adaptive clustering and multi-hop D2D communications.

The primary goal of this proposed model is to improve the coverage quality of service signals at the user devices in post-disaster scenarios. The multi-hop D2D acts to extend coverage services for user devices located outside of the BS coverage area.

As a result, the optimal relay hops are a promising technique for supporting B5G wireless communications by extending coverage areas where coverage is unavailable due to natural disasters. Thus, to enable effective communication and faster disaster recovery, optimal relays can change location dynamically in response to an emergency and have fast evictions. In post-disaster areas, the BS denotes the user device backhaul link.

*A. Relay hop selection*

In a post-disaster situation, the wireless coverage services are improved by selecting the optimal relay node to permit communication links between the BS and user devices in/out of the coverage area to reach remote user devices.

At the edge of coverage, relay nodes are chosen based on residual energy and reliable link quality between relay nodes in the coverage region and user devices in and out of the coverage area. This selection comes at the expense of complexity, as it requires selecting the optimal relay selection scheme by comparing a given transmit signal-to-noise ratio.

Thus, the BS acts as a centralized coverage service to relay nodes to maximize the received signal strength at the edge nodes. In addition, the BS selects which user nodes can act as relay nodes, connecting functioning and dysfunctional areas.

*B. Cluster head selection*

The resources of energy limitation are a problem that reduces wireless coverage service processing ability. Moreover, the cluster's head is responsible for forwarding its clients' cellular traffic to the BS (other users from the same cluster). Thus, the Cluster Heads act to improve connectivity with low power consumption.

Thus, the propagation of the channel from the BS to the relay nodes (BS-$R_i$) and the relay nodes to cluster heads ($R_i$-Cluster Heads) assume a prominent important role in evaluating energy efficiency performance for the 3-hop link for reliable connectivity in a post-disaster scenario. Hence, the CH selection with clustering and D2D-assisted links are utilized for sustainable connectivity, reducing power consumption, and enhancing the reliability and performance of network system coverage in disaster situations.

The model describes the stages of transferring the signal between the coverage and non-coverage (disaster and non-disaster areas) through the relay, with adaptive clustering and D2D communications. In Figure 1, the relay nodes ($R_i$) on the edge of active BS coverage receive the wireless covered single and forward it to the cluster heads (Cluster Heads) in the out-of-covered area. In addition, the Cluster Heads with residual energy more significant than the threshold will establish the communication link with cluster members (CMs) as D2D communication within the cluster. Then, the signal to interference plus noise ratio (SINR) at a receiver in the system's $i^{th}$ user device as a relay, Cluster Heads, and multi-hop D2D communication is denoted as follows:

$$SINR_i = \frac{p_i h_i}{\sum_k^n p_k h_k - p_i h_i + \sigma^2} \quad (1)$$

where $p_i$ is the transmission power of the $BS$, n is the number of user devices, $h_i$ is the attenuation from the $i^{th}$ user device to the BS to the relay node $R_i$, and $\sigma^2$ is background noise.

According to [31], the capacity for route $i$ in the link $l$ of the relay, Cluster Heads and the D2D communication system is calculated as:

$$C_{i,l} = \sum_{i=1}^{N} B \log_2(1 + SINR_{i,l}) \quad (2)$$

where $B$ represents the communication bandwidth.

The overall instantaneous transmission vector for energy efficiency $(EE_{i,l})$ is the energy efficiency elements from every link L $i$,$l$, the system capacity for obtaining a large transmission scale of N-relay hops. In addition, it is considered a scalable technology for establishing a multi-hop communication link to extend the coverage area range.

Thus, the energy efficiency performance is denoted as:

$$EE_{i,l} = \frac{C_{i,l}}{H p_i} \quad (3)$$

where $H$ represents the number of hops on every link $i$, $l$, and $p_i$ represents the maximum transmission power.

TABLE II MATLAB SIMULATION PARAMETERS

| Symbol | Description | Value |
|---|---|---|
| $BStx$ | BS transmit power | 5 W |
| $Rtx$ | Relay transmit power | 2.5 W |

| Symbol | Description | Value |
|---|---|---|
| $CH_{tx}$ | Cluster head transmit power | 1.5 W |
| $d\text{-}hop_1$ | Active BS transmit distance | 100–1000 m |
| $d\text{-}hop_2$ | Relay transmit distance | 5–250 m |
| $d\text{-}hop_3$ | D2D transmit distance | 5–50 m |
| $H$ | Number of hops | 3 |
| $\alpha$ | Path loss exponent | 2, 2.5, 3 |
| $BW$ | Bandwidth | 10 MHz |
| $fc$ | Carrier frequency | 700 MHz |

## III. NUMERICAL RESULTS AND ANALYSIS

This section summarizes the simulation findings to determine the best relay performance for extending BS coverage services in post-disaster scenarios. The simulation's parameters are listed in Table II. The model is proposed to assist the PSN in disaster recovery through optimal relay and multi-hop D2D communication. The aim is to extend the BS coverage and provide wireless service to user devices in disaster areas.

### A. ENERGY EFFICIENCY IN HOP$_1$

Energy efficiency in hop1 represents the number of bits that can be sent over a unit of power consumption for the link between the BS and relay, and it is measured by bits per joule. Figure 2 shows the energy efficiency performance versus the distance between the active BS in the coverage area and the relay on the edge of the coverage area. In a disaster scenario, the traffic intensity increases due to the limited network resources and the increased probability of traffic loss due to limited resources. In this context, energy efficiency is a key important factor for communication during disasters to relieve time and power consumption for the relay hops. Therefore, we simulated the parameters that affect the increased energy efficiency during the relay hops. The performance and energy efficiency are stable at 0.36 Mbits/joule in the case of the distance of 100 m due to the fixed bandwidth and path loss. However, energy efficiency decreases with increasing the distance between the active BS and relay on the edge of coverage with different path loss exponents.

The energy efficiency performance decreases from 0.36 to 0.24 Mbits/joule in the case of the BS–relay distance increasing from 100 m to 1000 m and $\alpha$ = 2.1. Energy efficiency decreases from 0.36 to 0.27 Mbits/joule at $\alpha$ = 2.3. However, the energy efficiency decreases from 0.36 to 0.29 Mbits/joule at $\alpha$ = 2.8 due to the decreased spectral efficiency for the channels and the increased path loss of the received signals at the nodes.

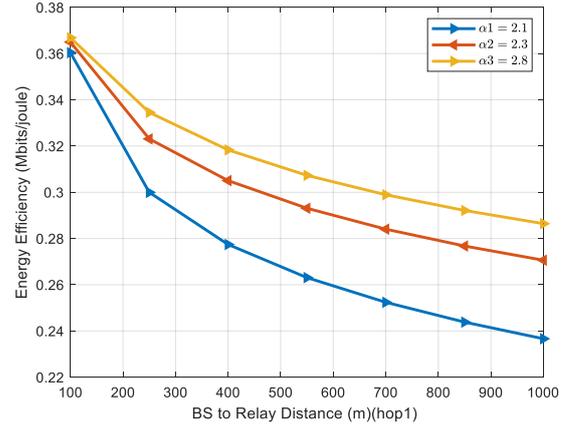

Figure 2: Energy efficiency performance versus distance between the active BS and relay with different path loss exponents and bandwidths.

### B. ENERGY EFFICIENCY IN HOP$_2$

Figure 3 shows the energy efficiency versus the relay–CH distance for communication during disasters. The performance and energy efficiency are stable between 225 to 230 Mbits/joule in the case of the distance of 100 m due to the fixed bandwidth and path loss. The energy efficiency performance decreases for each case of $\alpha$ = 2.1, 2.3, 2.8 and the effective relay–CH distance from 100 to 250 m. However, it can be seen that the performance of the energy efficiency decreases from 225 to 186 Mbits/joule at $\alpha$ = 2.1, from 230 to 203 Mbits/joule at $\alpha$ = 2.3, and from 230 to 210 Mbits/joule at $\alpha$ = 2.8 when the distance increases from 100 to 250 m due to the decrease in spectral efficiency for the channels and the increased path loss of the received signals at the nodes.

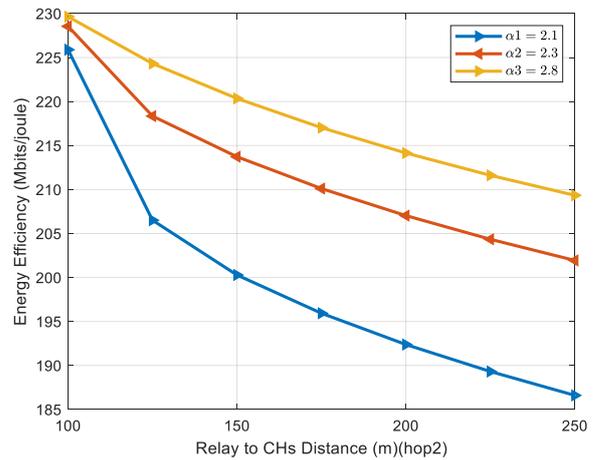

Figure 3: Energy efficiency performance versus distance between the active relay and Cluster Heads with different path loss exponents.

### C. ENERGY EFFICIENCY IN HOP$_3$

Figure 4 shows the analysis of energy efficiency versus D2D distance with different $\alpha$. It can be seen that the energy efficiency decreases in the low distance between D2D communications around less than 50 m due to co-channel interferences for D2D that affect the system's performance. The energy efficiency decreases when the distance increases from 5 m to 50 m due to high interference between the D2D communication that enables saving power for the devices.

D. ENERGY EFFICIENCY PERFORMANCE IN THREE HOPS

Figure 5 shows the performance of the three hops in terms of energy efficiency. The energy efficiency performance of the $hop_1$ between the active BS to the relay nodes can be seen to be maximized at 490 Mbits/joule. The BS's fixed power supply provides continuous energy for the connectivity in this $hop_1$. The energy efficiency performance in the second $hop_2$ between the relay hops and Cluster Heads decreases to 280 Mbits/joule. This is due to the propagation channel path losses caused by the loss of sent coverage signals with distance. Moreover, the $hop_3$ comes with low energy efficiency due to the short distance between the D2D communication and the increased interference that affects the coverage services.

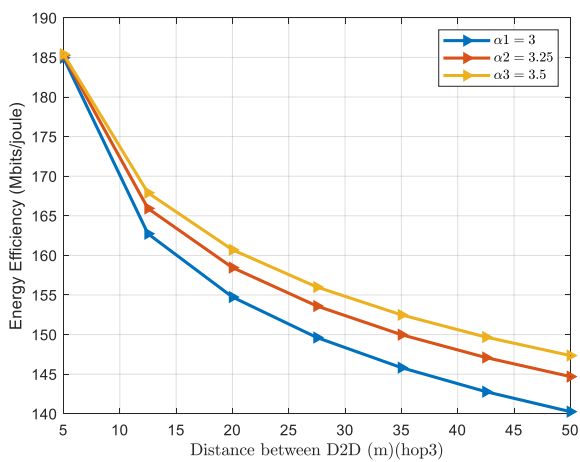

Figure 4: Energy efficiency performance versus distance between the CMs and CMs (D2D communication) with different path loss exponents.

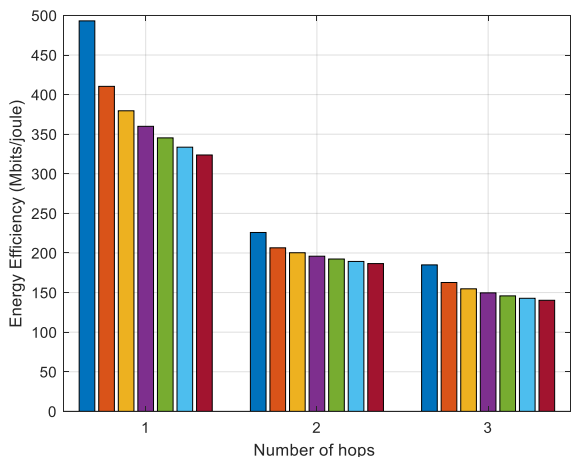

Figure 5: Comparison of performance of energy efficiency for three hops.

Figure 6 shows the simulation result in which we analyzed the energy efficiency of the proposed model and the benchmark performance in [19] of the D2D multi-hop communication with various distance settings. The energy efficiency performance decreases with increased D2D communication distances due to increased path loss between the source and distant nodes that affect received signal strength. Therefore, the energy efficiency performance for the proposed model decreases from 620 to 491 Mbits/joule at 10 MHz bandwidth. However, the energy efficiency performance for the model in [19] decreased from 620 to 467 Mbits/joule. Results show the energy efficiency of multi-hop D2D communications is enhanced to extend out-of-coverage service by around 300 m for support scenarios for disaster network recovery. The edge coverage area's relay hops can also manage the BS's functioning and transfer coverage service to the disaster area. Additionally, our suggested network performs better in terms of energy economy than the model in [16], which improves lifetime connectivity in disaster recovery.

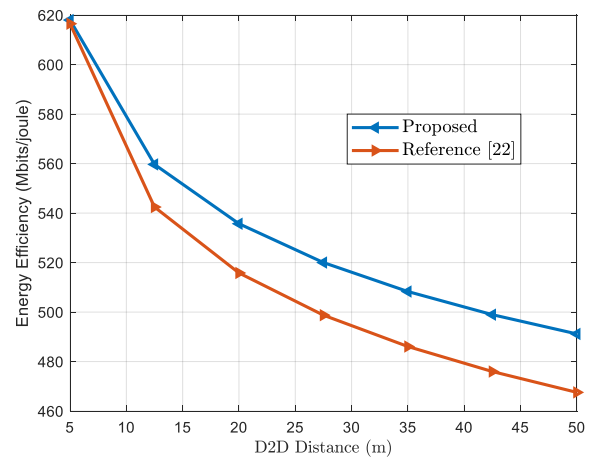

Figure 6: The proposed and benchmark models compare energy efficiency performance versus D2D distance.

IV. CONCLUSIONS

B5G is gaining so much interest from industry, government, and academia that adequate arrangements for post-disaster EMS are required. Furthermore, the specific nature of a post-disaster network situation is unknown; nonetheless, the notion of autonomous and resilient systems should be considered. This paper finds improving the BS coverage service through the hop's communications between the in-coverage and out-of-coverage areas. The energy efficiency performance of the three hops was analyzed based on different path loss exponents and bandwidths for three hops upon densities to evaluate coverage in a disaster scenario. Establishing D2D multi-hop communication is also crucial in disaster scenarios, as it increases the energy efficiency to prolong network connectivity.